\documentclass[journal]{IEEEtran}

\usepackage{cite}
\usepackage{amssymb}
\usepackage{algorithm}
\usepackage{algorithmic}
\usepackage{epsfig}
\usepackage{subfigure}
\usepackage{multirow}
\usepackage{float}
\usepackage{amsmath}
\usepackage{longtable}
\usepackage{booktabs}
\usepackage{color}
\usepackage{subfigure}
\usepackage{fancyhdr}
\usepackage{epstopdf}
\usepackage{bm}
\usepackage{amsopn}
\usepackage{lineno}

\begin{document}

\title{Quasi Time-Fuel Optimal Control Strategy for dynamic target tracking   }
\author{Huaihang Zheng, Junzheng Wang, Dawei Shi, \IEEEmembership{Senior Member, IEEE}, Dongchen Liu, Shoukun Wang
\thanks{This paper is submitted for review on April 19, 2021. This work was supported in part by the National Key Research and Development Project of China under Grant 2019YFC1511401 and the National Natural Science Foundation of China under Grant 51675041 and Grant 61773060. (Corresponding author: Junzheng Wang.) 
 }
\thanks{H. Zheng, J. Wang, D. Shi, D. Liu, and S. Wang are with the School of Automation, Beijing Institute of Technology, Beijing 100081, China (e-mails: hhzheng@bit.edu.cn, wangjz@bit.edu.cn, daweishi@bit.edu.cn, liudongchen@bit.edu.cn, bitwsk@bit.edu.cn).
}
}

\maketitle

\begin{abstract}
A quasi time-fuel optimal control strategy to solve the dynamic tracking problem of unmanned systems is investigated. It could bring the biggest advantage into full play in the field of switching tracking for multiple targets, such as the multi-target strike of weapons and the rapid multi-target grabbing of robots on industrial assembly lines. Compared with the time-fuel optimal control studied before, the proposed controller retains the advantages of optimal control when switching between multiple dynamic targets, and overcomes the high-frequency oscillation problem of the system under the discontinuous control strategies. Moreover, the asymmetry of friction load, which can affect the dynamic performance of the system, is also considered. Therefore, the novel control law proposed in this brief can make the corresponding system achieve the desired transient performance and satisfactory steady-state performance when switching between multiple dynamic targets. The experiment results based on the visual tracking turntable verify the superiority of the proposed method.
\end{abstract}


\section{Introduction}
\label{sec:introduction}
\IEEEPARstart{M}{otion} control and motion planning are closely related in the dynamic tracking problem of unmanned systems \cite{zhang2018compatible, huang2019motion, mccarragher1994petri}. In some studies, motion control and motion planning will be fused, which is the core research topic of unmanned control systems represented by robots \cite{ma20173}. It will become particularly challenging, especially when the unmanned control system is limited by actual factors such as mechanical structure, economic cost and equipment size.
Many researchers have made contributions to the motion planning problem under the control constraints of unmanned systems \cite{2009Hybrid,sun2017disturbance, sun2019novel}. These control algorithms enable unmanned systems to avoid the adverse effects of control saturation. 
Besides the transient response of the system in the control process, which has been widely investigated, how to adjust control strategies according to environmental conditions should also be taken into consideration, so that unmanned systems can obtain longer working hours with sufficient response speed.
That is, the control strategy designed for the unmanned system needs to find a balance among multiple performance indices.

For problems with mixed performance indices and control constraints, optimal control theory provides a good research direction. In \cite{cao2012fuel}, the optimal mixed depletion mode control strategies were constructed and obtained by reasonable and balanced using of the battery power and the engine power. An in-vehicle energy optimal control (EOC) improving the efficiency of the motor drive system was proposed in \cite{zhang2019energy} to solve the problem of short driving distance per charge for electric vehicles.
Meanwhile, there have been numerous efforts to explore the energy-saving or fuel-saving control strategies for unmanned systems, and remarkable success was achieved using Pontryagin's minimum principle \cite{Low-Thrust,kim2010optimal, chowdhury2020optimal, yang2017optimizing}. Among them, Singh \emph{et al.} \cite{Low-Thrust}  investigated the manifolds of three Near-Rectilinear Halo Orbits (NRHOs) and optimal low-thrust transfer trajectories using a high-fidelity dynamical model and the relative merits of the stable/unstable manifolds are studied with regard to time- and fuel-optimality criteria, for a set of representative low-thrust family of transfers.
Pioneered by Bobrow \emph{et al.} \cite{bobrow1985time}, the time optimal trajectory planning of industrial manipulator was investigated. Recently, time-dependent optimal control problem has been widely explored in many domains, including spacecraft control systems \cite{kim2014near,2018AAS},  robot control systems \cite{ he2020time}, and the power system with model uncertainty \cite{ zhang2019high}.
Furthermore, a hybrid optimal control problem is formulated and solved in the context of efficient piecewise optimal transfers to the lunar gateway by pre-computing the terminal coast arcs \cite{2021Eclipse}. It makes the problem much easier to converge especially in the presence of any external perturbations as shown in the paper.

According to the research on unmanned systems mentioned above, we found that the unmanned systems usually need to balance rapidity and economy in the control process. For example, an unmanned platform with limited fuel requires a longer standby time when it is cruising, and requires a faster response speed when it is tracking a target.
This type of compromise between the shortest time control and the most fuel-efficient control is named as the time-fuel optimal control problem \cite{2003A}.
In the 1960s, time-fuel optimal feedback control law of the ideal double integrator were systematically derived by Athans and Falb \cite{athans2013optimal}. 
 However, if the control strategies proposed in \cite{2003A} is applied to the perturbed double integrator system, there may be some undesirable phenomena (i.e., chattering and overshoot \cite{bang1999feedback,sidi1997spacecraft,forni2010family}), since the time-fuel optimal control systems are sensitive to disturbance, unmodeled dynamics, and parameter variations \cite{1993Proximate}. 
 The most popular alternative approach is the so-called proximate time-optimal servomechanism (PTOS) \cite{2012Improved}. This approach starts with a near-time-optimal controller, and then switches to a linear controller when the system output is close to a given target \cite{workman1987adaptive}. 
 What’s more, to overcome chattering and overshoot, \cite{jing2002memorised} proposed an improved control law which introduced which introduced two compensation factors of perturbations into switching functions and added memory to control law.

This brief discusses the problem of the trade-off between time and fuel consumption when the double integrator system tracks dynamic targets. Moreover, this brief solves the highly oscillatory behavior caused by the imperfection of the actual system through appropriate improvements when tracking dynamic targets, and then proposes a quasi time-fuel optimal control strategy (QTFOC). This algorithm could bring the biggest advantage into full play in the field of multi-target switching tracking, such as the multi-target strike of weapons and the rapid multi-target grabbing of robots on industrial assembly lines. Main contributions and novelties of this brief can be summarized as follows.

\begin{enumerate}
	\item  The quasi time-fuel optimal control strategy is proposed for tracking dynamic targets based on Pontryagin's minimum principle. 
	It extends the previously studied time-fuel optimal control strategy from approaching static targets to tracking dynamic targets, and provides an important basis for unmanned double integrator systems to balance working periods and response speed.  
	\item Local linear feedback and two nonlinear buffers are introduced to solve the oscillatory behavior caused by high frequency switching of control input when quasi time-fuel optimal control strategy is applied to an actual system. The essence of this improvement is to increase a small amount of cost in exchange for better robustness of the control strategy.
	\item The ultimate performance of the actual system is further explored by taking the asymmetry of friction load into consideration. 
	\item Experimental verification is conducted using a visual tracking system. The results have demonstrated the effectiveness and feasibility of the proposed quasi time-fuel optimal control strategy.
\end{enumerate}

The remainder of this brief is organized as follows. In Section II, the preliminary and problem formulation are provided. In Section III, we present a quasi time-fuel optimal control strategy. In Section IV, the influence the asymmetry of the friction load on the system is analyzed. In Section V, we carry out the experiments based on the visual tracking turntable driven by servo motor. The conclusions are given in Section VI.

\section{Preliminaries }

\begin{figure}[htbp]
\centerline{\includegraphics[ width=0.9\columnwidth]{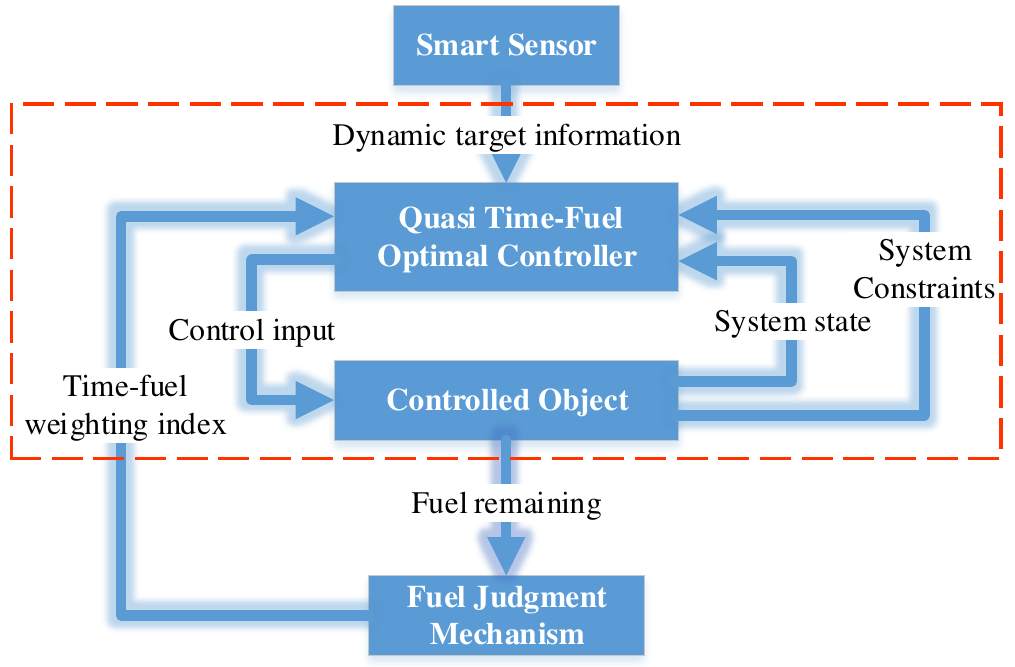}}
\caption{Block diagram of the overall system.}
\label{FIG_2-1}
\end{figure}

We investigate a visual tracking system in the form of a double integrator, which includes smart sensors, the fuel judgment mechanism and the quasi time-fuel optimal controller (see \mbox{Fig. \ref{FIG_2-1}}). The red dashed box in \mbox{Fig. \ref{FIG_2-1}} is the focus of this brief. For this constrained and fuel-limited double integrator system, the system constrains, transient response, and fuel consumption are all need to be taken into consideration for better control performance. The model of the system is given by
\begin{align}
    \dot{x}_{1}&=x_{2},\nonumber\\
\dot{x}_{2}&=u, \quad \left |u \right |\leq M.
\label{94-2}
\end{align}

Therefore, the performance index can be defined as
\begin{align}
    J(u)=\int_{t_{0}}^{t_{f}}\left [ \lambda +\left | u(t)\right |\right ]dt,
\label{2-10}
\end{align}
where  $t_{0}$ and $t_{f}$ represent the initial time and terminal time respectively. $\lambda$ indicates the weight of time and $\left \{\lambda \in \mathbb{R}\mid \lambda >0\right \}$. 

According to Pontryagin's minimum principle, the time-fuel optimal control for the double integrator system is as follow \cite{athans2013optimal}:

\begin{align}
    u(t)=\left\{\begin{matrix}
    +M, & p_{2}(t)< -1,\\ 
     0,& -1<p_{2}(t)< 1,\\ 
     -M,& p_{2}(t)>1,\\ 
     -v(t)*{\rm sign}[p_{2}(t)], & \left | p_{2}(t)\right |=1,
    \end{matrix}\right.
\label{2-11}
\end{align}
where $p^{{\rm T}}(t)=[p_{1}(t),p_{2}(t)]$ is the costate of the system. The optimality condition stipulates that the time derivative of costate $p^{{\rm T}}(t)$ satisfies:
\begin{equation}
\left\{
\begin{array}{l}
\dot{p}_{1}(t) = 0,\\ 
 \dot{p}_{2}(t) = -p_{1}(t),
\end{array}
\right.
\label{2-12}
\end{equation}
and
\begin{equation}
\left\{
\begin{array}{l}
p_{1}(t)=c_{1},  \\ 
 p_{2}(t)=-c_{1}t+c_{2}.
\end{array}
\right.
\label{2-13}
\end{equation}

\textit{Definition 1:} If the tracking error between the system and the dynamic target under the control strategy is bounded, and the tracking error will be reduced to 0 when the dynamic target speed $x_{2d}(t_{c})=0$, where $t_{c}$ represents the time of each control moment, then the system is said to be able to keep up with the dynamic target.

\textit{Remark 1:} Since $p_{2}(t)$ is a linear function of time $t$, the possible optimal control input sequences include: $\left \{0\right \}$, $\left \{+M\right \}$, $\left \{-M\right \}$, $\left \{0,+M\right \}$, $\left \{0,-M\right \}$, $\left \{+M,0\right \}$, $\left \{-M,0\right \}$, $\left \{+M,0,-M\right \}$, $\left \{-M,0,+M\right \}$. Moreover, when $u(t)$ happens to switch from $+M$ or $0$ to $0$ or $+M$, $p_{2}(t)=1$. Similarly, when $u(t)$ happens to switch from  $-M$ or $0$ to $0$ or $-M$, $p_{2}(t)=-1$. In the light of Definition 1, the system described in this brief can maintain its current state to save fuel when the error between the system and the target is bounded and constant. Therefore, the control sequences $\left \{+M,0\right \}$ and $\left \{-M,0\right \}$ are feasible and caused by $x_{2d}(t_{c})\neq 0$.

It is noteworthy that $x_{1d}(t)$ and $x_{2d}(t_{c})$ represent the target position and speed of the system respectively and the future motion state of the target is completely unknown to our system, and sometimes the target state will suddenly change due to the switching of the target. In order to track the dynamic target, the current motion state of the target will be updated at each control moment through additional sensors. This indicates that ${{x}_{1d}}(t)$ is related to ${{x}_{2d}}({{t}_{c}})$ and will be updated in real time, while ${{x}_{2d}}({{t}_{c}})$ can only be updated through the real-time feedback from the sensors.

Based on the above conditions, the following parts will explain how to get the switch plane of time-fuel optimal control under the dynamic target step by step, and further improvements are proposed to solve the problems of asymmetrical friction load and system oscillation. Finally, a very practical control strategy is obtained.

\section{Preliminary Quasi Time-fuel Optimal Control Strategy for Dynamic target Tracking}
In this section, we explore the analytical solution of quasi time-fuel optimal control strategy based on dynamic terminal target set. \cite{bang1999feedback,sidi1997spacecraft,forni2010family} showed that the application of time-optimal feedback control law to the perturbed double integrator gives rise to undesired phenomena, such as chattering and overshoot. For the double integrator system in this brief, the time-fuel optimal control strategy (TFOC) will also induce a highly oscillatory behavior of the system near the target. Therefore, it is necessary to 
use local linear feedback and some buffer areas to overcome the chattering problem caused by model uncertainties or feedback delays, and optimize the convergence performance of the system.

\begin{figure}[htbp]
\centerline{\includegraphics[width=0.7\columnwidth]{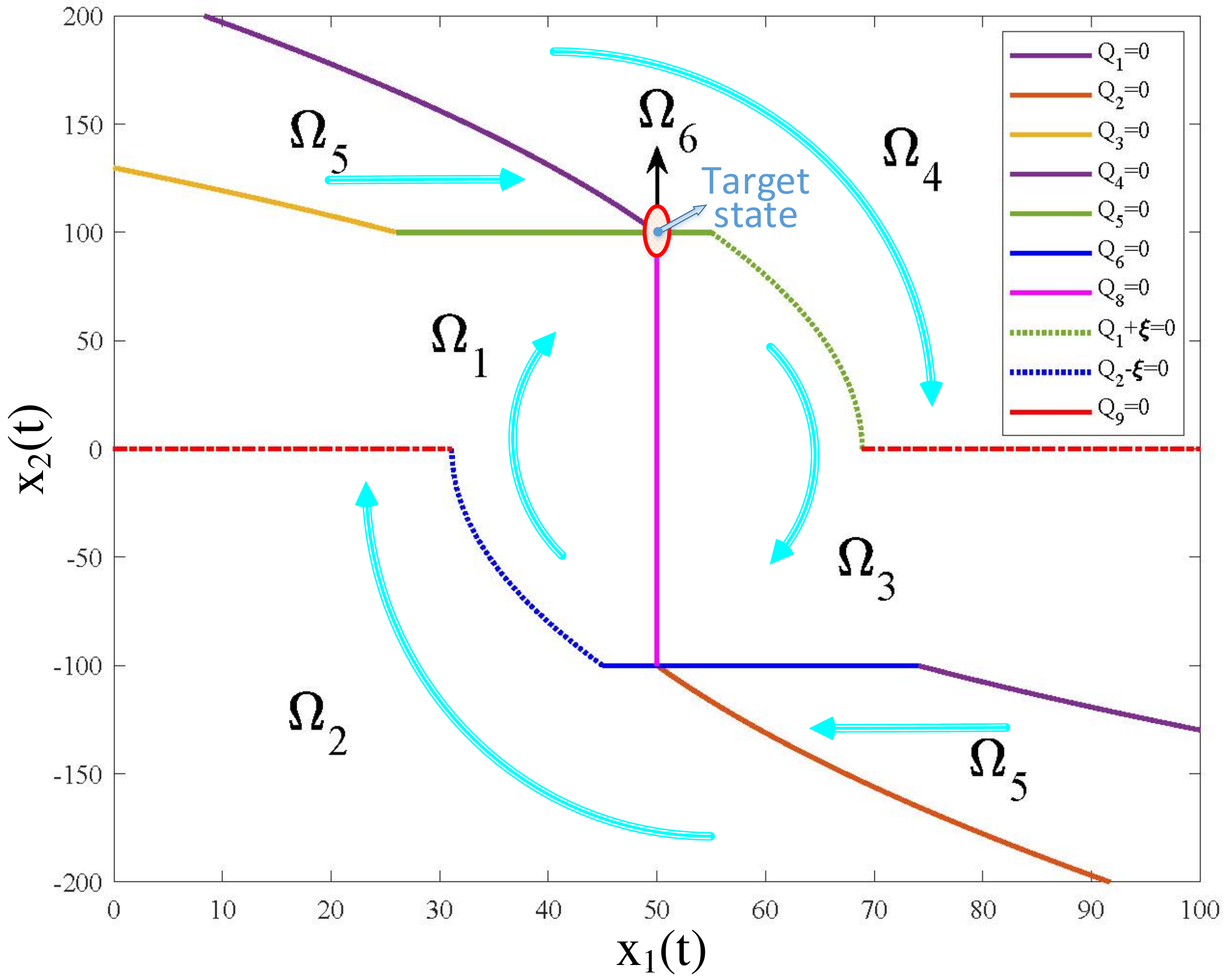}}
\caption{Switch plane of the control strategy (QTFOC).}
\label{FIG_5}
\end{figure}

Thus, a practical quasi time-fuel optimal control strategy for the double integrator system is:

\begin{enumerate}
    \item
    $u(t)=M$, iff $(x_{1},x_{2} )\in\Omega_{1}$, where
    \begin{align}
\Omega_{1}&=\left \{(x_{1},x_{2})\mid Q_{9}> 0\cap  \left \{\Omega_{11}\cup\left \{\Omega_{12}\cap \Omega_{13}\right \}\right \}\right \}.
\label{93-6}
     \end{align}
The subset of $\Omega_{1}$ is defined as follows:   
\begin{equation}
\left\{
\begin{array}{l}
\Omega_{11}=\left \{(x_{1},x_{2})\mid Q_{5} \geq 0\cap Q_{3}\leq 0\right \}\\ 
 \Omega_{12}=\left \{(x_{1},x_{2})\mid Q_{5}\leq 0 \cap Q_{8}\leq 0\right \}\\ 
\Omega_{13}=\left \{(x_{1},x_{2})\mid (Q_{2}+\xi\geq0)\cap Q_{6}\geq 0\cup x_{2}\geq 0
\right \}\nonumber
\end{array}
\right.
\end{equation}

	\item  
$u(t)=Mmin\left \{1,\left | \frac{Q_{2}}{\xi}\right |\right \}$,  iff $(x_{1},x_{2} )\in\Omega_{2}$, where
\begin{align}
\Omega_{2}=\left \{(x_{1},x_{2})\mid Q_{9}> 0\cap \left \{\Omega_{21}\cup \Omega_{22}\right \}\right \}.
\label{93-15}
\end{align}
The subset of $\Omega_{2}$ is defined as follows:   
\begin{equation}
\left\{
\begin{array}{l}
\Omega_{21}= \left \{(x_{1},x_{2})\mid Q_{2}\leq 0\cap Q_{6}<0 \cap(Q_{2}+\xi)\geq 0\right \}\\
\Omega_{22}=\left \{(x_{1},x_{2})\mid (Q_{2}+\xi)<0\cap x_{2}<0\right \}\nonumber
\end{array}
\right.
\end{equation}

 \item
    $u(t)=-M$, iff $(x_{1},x_{2} )\in\Omega_{3}$, where
    \begin{align}
\Omega_{3}&=\left \{(x_{1},x_{2})\mid Q_{9}> 0\cap  \left \{\Omega_{31}\cup\left \{\Omega_{32}\cap \Omega_{33}\right \}\right \}\right \}.
     \end{align}

The subset of $\Omega_{3}$ is defined as follows: 
\begin{equation}
\left\{
\begin{array}{l}
\Omega_{31}=\left \{(x_{1},x_{2})\mid Q_{6} \leq 0\cap Q_{4}\geq 0\right \}\\ 
 \Omega_{32}=\left \{(x_{1},x_{2})\mid Q_{6}\geq 0 \cap Q_{8}\geq 0\right \}\\ 
\Omega_{33}=\left \{(x_{1},x_{2})\mid (Q_{1}-\xi\leq0)\cap Q_{5}\leq 0\cup x_{2}\leq 0\right \}\nonumber
\end{array}
\right.
\end{equation}
	\item 
$u(t)=-Mmin\left \{1,\left | \frac{Q_{1}}{\xi}\right |\right \}$,  iff $(x_{1},x_{2} )\in\Omega_{4}$, where
\begin{align}
\Omega_{4}=\left \{(x_{1},x_{2})\mid Q_{9}> 0\cap \left \{\Omega_{41}\cup \Omega_{42}\right \}\right \}.
\label{93-17}
\end{align}
The subset of $\Omega_{4}$ is defined as follows:  
\begin{equation}
\left\{
\begin{array}{l}
\Omega_{41}= \left \{(x_{1},x_{2})\mid Q_{1}\geq 0\cap Q_{5}>0 \cap(Q_{1}-\xi)\leq 0\right \}\\
\Omega_{42}=\left \{(x_{1},x_{2})\mid (Q_{1}-\xi)>0\cap x_{2}>0\right \}\nonumber
\end{array}
\right.
\end{equation}
	\item 
$u(t)=0$, iff $(x_{1},x_{2} )\in\Omega_{5}$, where 
\begin{align}
\Omega_{5}=\left \{(x_{1},x_{2})\mid Q_{9}>0 \cap \left \{\Omega_{51}\cup \Omega_{52}\right \}\right \}.
\label{93-18}
\end{align}

The subset of $\Omega_{5}$ is defined as follows:   \begin{align}
\left\{\begin{matrix}
\Omega_{51}=\left \{(x_{1},x_{2})\mid Q_{1}<0 \cap Q_{3}> 0\cap Q_{5}> 0\right \}\\ 
\Omega_{52}=\left \{(x_{1},x_{2})\mid Q_{2}>0\cap Q_{4}< 0\cap Q_{6}< 0\right \}\nonumber
\end{matrix}\right.
\end{align}

     \item 
     $  u=-M\left [(x_{1}(t)-x_{1d}(t))+\frac{2}{\sqrt{M}}(x_{2}(t)-x_{2d}(t_{c}))\right ]$, iff $(x_{1},x_{2} )\in\Omega_{6}$, where
 \begin{align} 
  \Omega_{6}=\left \{(x_{1},x_{2})\mid Q_{9}\leq 0\right \}.
  \label{93-19}
 \end{align} 
 \end{enumerate}
Note that $\delta \geq 0$, and $\xi$ is the width of the buffer areas which are used to eliminate the highly oscillatory behavior of the system. The different regions ($\Omega_{1}$ to $\Omega_{6}$) of $\mathbb{R}^2$ labeled in \mbox{Fig. \ref{FIG_5}} are divided by switch curves ($Q_{1}$ to $Q_{9}$) and the switch curves are described mathematically in \mbox{Table \ref{table_01}}. Without loss of generality, the value of $\delta$ is set to 0 in \mbox{Fig. \ref{FIG_5}} and the light blue line with arrow indicates the direction of the system state change. 

The local linear control region $\Omega_{6}$ is determined by $Q_{9}\leq 0$. It is a ellipse with the following parameters:
\begin{align}
k_{F}&=\frac{Ma_{4}(a_{1}-\sqrt{M}a_{2}+\frac{Ma_{4}}{4})-(\sqrt{M}a_{2}-\frac{Ma_{4}}{2})^{2}}{4(a_{1}-\sqrt{M}a_{2}+\frac{Ma_{4}}{4})},\nonumber\\
a_{1}&=\frac{M+5}{4\sqrt{M}},\quad
a_{2}=\frac{1}{2M},\quad
a_{4}=\frac{1+M}{4M\sqrt{M}}.\nonumber
\end{align}
In the Appendix I, we will show that 1) there exists a Lyapunov function $\boldsymbol{V}=\boldsymbol{x_{e}}^{\rm T}\boldsymbol{P}\boldsymbol{x_{e}}$ for the system when $(x_{1},x_{2} )\in\Omega_{6}$ which has a negative definite derivative when the state is on the boundary of $\Omega_{6}$; and hence, the system is stable to the region of $\Omega_{6}$. 2) Any trajectory starting outside $\Omega_{6}$ will enter $\Omega_{6}$.

\begin{table}[htbp]
\renewcommand\arraystretch{1.3}
\centering
\caption{Mathematical description of the swtich curves}
\label{table_01}
\setlength{\tabcolsep}{3pt}
\begin{tabular}{|p{25pt}|p{170pt}|}
\hline
Symbol& 
Mathematical Description \\
\hline
$ Q_{1} $& 
$x_{1}(t)+\frac{1}{2M}x_{2}^{2}(t)-x_{1d}(t)-\frac{1}{2M}x_{2d}^{2}(t_{c})$ \\
$ Q_{2} $& 
$x_{1}(t)-\frac{1}{2M}x_{2}^{2}(t)-x_{1d}(t)+\frac{1}{2M}x_{2d}^{2}(t_{c})$ \\
$ Q_{3} $& 
$x_{1}(t)+\frac{\lambda +4M}{2M\lambda }x_{2}^{2}(t)-x_{1d}(t)-\frac{1}{2M}x_{2d}^{2}(t_{c})$ \\
$ Q_{4} $& 
$x_{1}(t)-\frac{\lambda +4M}{2M\lambda }x_{2}^{2}(t)-x_{1d}(t)+\frac{1}{2M}x_{2d}^{2}(t_{c})$ \\
$ Q_{5} $& 
$x_{2}(t)-\left |x_{2d}(t_{c}) \right |-\delta$ \\
$ Q_{6} $& 
$x_{2}(t)+\left |x_{2d}(t_{c}) \right |+\delta$ \\
$ Q_{7} $& 
$x_{2}(t)-x_{2d}(t_{c})$ \\
$ Q_{8} $& 
$x_{1}(t)-x_{1d}(t)$ \\
$ Q_{9} $& 
$a_{1}(x_{1}(t)-x_{1d}(t))^{2} +a_{4}(x_{2}(t)-x_{2d}(t_{c}))^{2}+2a_{2}(x_{1}(t)-x_{1d}(t))(x_{2}(t)-x_{2d}(t_{c}))-k_{F}$ \\

\hline
\end{tabular}
\label{tab01}
\end{table}

In order to make the description more concise, we will explain how the boundaries of other regions are obtained without considering the local linear region. The $\Omega_{5}$ is very critical in this biref, so we start the analysis of QTFOC from this region where $u(t)=0$.

Now we need to do some preparatory work for the analysis of $\Omega_{5}$. The control input sequence in Remark 1 will be discussed separately to obtain the switching boundary of $\Omega_{5}$. The initial state of the system is set to $x_{0}=[x_{10},x_{20}]$.

\textit{Definition 2:} Suppose there is a control input sequence $u_{n}(t)=\left \{\phi_{1},\phi_{2},\cdots ,\phi_{i},\phi_{i+1},\cdots ,\phi_{n-1},\phi_{n} \right \}$, $n\in \mathbb{N}+$, $i\in (1,2,\cdots ,n)$, then $u(t)=\left \{\phi_{i},\phi_{i+1},\cdots ,\phi_{n-1},\phi_{n} \right \}$ is called a control input sub-sequence of $u_{n}(t)$.

Note that the control input of the time-fuel optimal control problem is only related to the current state of the system and the target. Therefore, $u_{1}(t)$ is a control input sub-sequence of $u_{2}(t)$, which means that $u_{1}(t)$ can be regarded as a special case of $u_{2}(t)$. According to the possible optimal control input sequences of the system discussed in Remark 1, it is obvious that $\left \{0\right \}$ is the control input sub-sequence of $\left \{+M,0\right \}$ and $\left \{-M,0\right \}$. $\left \{+M\right \}$ and $\left \{0,+M\right \}$ are the control input sub-sequences of $\left \{-M,0,+M\right \}$. $\left \{-M\right \}$ and $\left \{0,-M\right \}$ are the control input sub-sequences of $\left \{+M,0,-M\right \}$. So for the optimal control input sequences in Remark 1, we only need to study $\left \{+M,0\right \}$, $\left \{-M,0\right \}$, $\left \{-M,0,+M\right \}$, $\left \{-M,0,+M\right \}$.

It is obvious that $x_{1}(t)=\frac{1}{2M}x_{2}^{2}(t)+c_{5}$ will be satisfied in $[t_{a},t_{b}]$ if $u(t)=+M$, and $x_{1}(t)=-\frac{1}{2M}x_{2}^{2}(t)+c_{6}$ will be satisfied in $[t_{c},t_{d}]$ if $u(t)=-M$, where $c_{5}\in \mathbb{R}$, $c_{6}\in \mathbb{R}$ \cite{2003Optimal}. Then, we analyze the situation of $u(t)=\left \{+M,0,-M\right \}$ and  $u(t)=\left \{-M,0,+M\right \}$. Suppose the current state of the dynamic target is $[x_{1d}(t),x_{2d}(t_{c})]$. When $u(t)=\left \{+M,0,-M\right \}$, we assume that the switching instants are $t_{1}^{-}$ and $t_{2}^{-}$. 
Obviously, $x_{2}(t_{1}^{-})=x_{2}(t_{2}^{-})>x_{2d}(t_{c})$ need to be satisfied in this situation and the system state at $t_{2}^{-}$ will satisfy the following equation:
\begin{align}
     x_{1}(t_{2}^{-})=-\frac{1}{2M}x_{2}^{2}(t_{2}^{-})+x_{1d}(t_{2}^{-})+\frac{1}{2M}x_{2d}^{2}(t_{c}).
\label{3-16}
\end{align}
It is noteworthy that the form of switch curve $Q_{1}$ and \eqref{3-16} is the same. It is called the switching boundary function from $u(t)=0$ to $u(t)=-M$ in this situation. During $[t_{1}^{-},t_{2}^{-}]$, $u(t)=0$ so that 
\begin{equation}
\left\{
\begin{array}{l}
\dot{x}_{1}(t)=x_{2}(t),\\ 
\dot{x}_{2}(t)=0.
\end{array}
\right.
\end{equation}
Integrate both sides of the above formulas:
\begin{equation}
\left\{
\begin{array}{l}
 x_{2}(t_{2}^{-})=x_{2}(t_{1}^{-}),\\ 
x_{1}(t_{2}^{-})=x_{1}(t_{1}^{-})+x_{2}(t_{1}^{-})\left [t_{2}^{-}-t_{1}^{-} \right ].
\label{3-18}
\end{array}
\right.
\end{equation}
Substituting \eqref{3-18} into \eqref{3-16} yields:
\begin{align}
    x_{1}(t_{1}^{-})+x_{2}(t_{1}^{-})\left [t_{2}^{-}-t_{1}^{-} \right ]+\frac{1}{2M}x^{2}_{2}(t_{1}^{-})\nonumber \\
    =x_{1d}(t_{1}^{-})+\frac{1}{2M}x_{2d}^{2}(t_{c}).
\label{3-19}
\end{align}
In order to obtain the relationship between $c_{1}$ and $c_{2}$, we consider the singular values of $p_{2}(t)$, i.e., $p_{2}(t)=\pm 1$. Therefore,
\begin{equation}
\left\{
\begin{array}{l}
-c_{1}t_{1}^{-}+c_{2}=-1,\\ 
-c_{1}t_{2}^{-}+c_{2}=1.
\end{array}
\right.
\end{equation}
Moreover, the following formula can be obtained:
\begin{align}
    \left [ t_{2}^{-}-t_{1}^{-}\right ]=-\frac{2}{c_{1}}.
\end{align}
For a steady state system, the Hamiltonian equals to zero, which is:
\begin{align}
    0=\lambda +p_{1}(t)x_{2}(t)+p_{2}(t)u(t)=\lambda+c_{1}x_{2}(t_{1}^{-}).
\end{align}
In this case, $x_{2}(t_{1}^{-})\neq 0$ so that
\begin{align}
[t_{2}^{-}-t_{1}^{-}]=\frac{2x_{2}(t_{1}^{-})}{\lambda }.
\label{3-23}
\end{align}
Substituting \eqref{3-23} into \eqref{3-19} yields:
\begin{align}
    x_{1}(t_{1}^{-})=-\frac{\lambda +4M}{2M\lambda }x_{2}^{2}(t_{1}^{-})+x_{1d}(t_{1}^{-})+\frac{1}{2M}x_{2d}^{2}(t_{c}).
\label{3-24}
\end{align}
It is noteworthy that the form of switch curve $Q_{3}$ and \eqref{3-24} is the same. The above formula is called the switching boundary function from $u(t)=+M$ to $u(t)=0$. 

\textit{Theorem 1:} If the control input sequence $\left \{+M,0,-M\right \}$ is applied to plant \eqref{94-2}, the following conditions will be satisfied:
\begin{equation}
\left\{
\begin{array}{l}
x_{1}(t_{1}^{-})< x_{1d}(t_{1}^{-})-\frac{2x_{2d}^{2}(t_{c})}{\lambda },\\ 
x_{2}(t_{1}^{-})> x_{2d}(t_{c}).
\end{array}
\right.
\label{93-1}
\end{equation}

\textit{Proof:}  When the control input sequence is $\left \{+M,0,-M\right \}$, it is obvious that $x_{2}(t_{1}^{-})=x_{2}(t_{2}^{-})>x_{2d}$. Refer to \eqref{3-24}, the constraint on $x_{1}(t_{1}^{-})$ is simply given by 
\begin{align}
x_{1}(t_{1}^{-})< x_{1d}(t_{1}^{-})-\frac{2x_{2d}^{2}(t_{c})}{\lambda }.
\end{align}
This completes the proof.\hfill $\Box$

When $u(t)=\left \{-M,0,+M\right \}$, we assume that the switching instants are $t_{1}^{+}$ and $t_{2}^{+}$. Obviously, $x_{2}(t_{1}^{+})=x_{2}(t_{2}^{+})< x_{2d}(t_{c})=x_{2d}(t_{c})$ need to be satisfied in this case. It is completely similar to $u(t)=\left \{+M,0,-M\right \}$ so that we can make the following analysis concisely. The switching boundary function from $u(t)=0$ to $u(t)=+M$ in this situation is that
\begin{align}
    x_{1}(t_{2}^{+})=\frac{1}{2M}x_{2}^{2}(t_{2}^{+})+x_{1d}(t_{2}^{+})-\frac{1}{2M}x_{2d}^{2}(t_{c}).
\end{align}
This is the switch curve $Q_{2}$. The switching boundary function from $u(t)=-M$ to $u(t)=0$ is that
\begin{align}
    x_{1}(t_{1}^{+})=+\frac{\lambda +4M}{2M\lambda }x_{2}^{2}(t_{1}^{+})+x_{1d}(t_{1}^{+})-\frac{1}{2M}x_{2d}^{2}(t_{c}),
\label{3-30}
\end{align}
which is the switch curve $Q_{4}$.

\textit{Theorem 2:} If the control input sequence $\left \{-M,0,+M\right \}$ is applied to plant \eqref{94-2}, the following conditions will be satisfied:
\begin{equation}
\left\{
\begin{array}{l}
x_{1}(t_{1}^{+})> x_{1d}(t_{1}^{+})+\frac{2x_{2d}^{2}(t_{c})}{\lambda },\\ 
x_{2}(t_{1}^{+})< x_{2d}(t_{c}).
\end{array}
\right.
\label{93-2}
\end{equation}

\textit{Proof:}   When the control input sequence is $\left \{-M,0,+M\right \}$, it is obvious that $x_{2}(t_{1}^{+})=x_{2}(t_{2}^{+})<x_{2d}(t_{c})=x_{2d}(t_{c})$. Refer to \eqref{3-30}, the constraint on $x_{1}(t_{1}^{+})$ is simply given by 
\begin{align}
x_{1}(t_{1}^{+})> x_{1d}(t_{1}^{+})+\frac{2x_{2d}^{2}(t_{c})}{\lambda }.
\end{align}
This completes the proof.\hfill $\Box$

Next, we will analyze the situation where the control input sequence is $\left \{+M,0\right \}$ and $\left \{-M,0\right \}$.

\textit{Lemma 1:} The optimal control sequence of plant \eqref{94-2} is $\left \{+M,0\right \}$ only if $x_{2d}(t_{c})>0$. The optimal control sequence of  plant \eqref{94-2} is $\left \{-M,0\right \}$ only if $x_{2d}(t_{c})<0$.

\textit{Proof:} The Hamiltonian of plant \eqref{94-2} with a performance index of \eqref{2-10} is
\begin{align}
    H=\lambda+\left | u(t)\right |+p_{1}(t)x_{2}(t)+p_{2}(t)u(t).
\end{align}
For a steady state system, when $u(t)=0$, the Hamiltonian will satisfies
\begin{align}
 H=\lambda+c_{1}x_{2}(t)=0,
\end{align}
because the system reaches the target state $\left [x_{1d}(t) \ x_{2d}(t_{c})\right ]^{{\rm T}}$ under the control of $u(t)=0$. Therefore, $c_{1}>0$ ($c_{1}<0$) when $x_{2d}(t_{c})<0$ ($x_{2d}(t_{c})>0$) since $\lambda>0$. On the other hand, from \eqref{2-11} and \eqref{2-13}, the control input sequence $u(t)=\left \{+M,0\right \}$ requires $p_{2}(t)$ to increase with time, which means $c_{1}<0$. The control input sequence $u(t)=\left \{-M,0\right \}$ requires $p_{2}(t)$ to decrease with time, which means $c_{1}>0$. This completes the proof.\hfill $\Box$

\textit{Remark 2:} According to Definition 1, the control input sequence $u(t)=\left \{\phi_{i},\phi_{i+1},\cdots ,\phi_{n-1}, 0  \right\}$ (i.e. $\phi_{n}=0$) allows the system to keep up with the dynamic target only if $x_{2d}(t_{c})\neq 0$ and
\begin{equation}
\nonumber
\left\{
\begin{array}{l}
x_{2}(t)= x_{2d}(t_{c})>0,\\ 
x_{1d}(t)>x_{10},
\end{array}
\right.
\quad {\rm or} \quad
\left\{
\begin{array}{l}
x_{2}(t)= x_{2d}(t_{c})<0,\\ 
x_{1d}(t)<x_{10}.
\end{array}
\right.
\end{equation}

\textit{Theorem 3:} If the control input sequence $\left \{+M,0\right \}$ is applied to plant \eqref{94-2}, there will be a switching instant $t_{1}^{-}$ in control process and the following conditions will be satisfied:
\begin{equation}
\left\{
\begin{array}{l}
x_{1}(t_{1}^{-})\geq x_{1d}(t_{1}^{-})-\frac{2x_{2d}^{2}(t_{c})}{\lambda },\\ 
x_{2}(t_{1}^{-})=x_{2d}(t_{c}).
\end{array}
\right.
\label{93-3}
\end{equation}

\textit{Proof:}  Referring to Lemma 1, it is obvious that the control input sequence $u(t)=\left \{+M,0\right \}$ means $x_{2d}(t_{c})>0$. Therefore,
\begin{align}
u(t)=\left\{\begin{matrix}\vspace{0.2cm}
+M, & t\in [0,t_{1}^{-}],\\ \vspace{0.2cm}
 0, & t\in [t_{1}^{-},\frac{x_{1d}(t_{1}^{-})-x_{1}(t_{1}^{-})}{x_{2d}(t_{c})}+t_{1}^{-}].
\end{matrix}\right.
\end{align}
Since the control input $u(t)$ of the system is equal to 0 before reaching the target state, we can obtain the following equation according to \eqref{94-2}:
\begin{equation}
\left\{
\begin{array}{l}
x_{1}(t)=c_{3}t+c_{4},\\ 
x_{2}(t)=c_{3},
\end{array}
\right.
\end{equation}
where $t\in [t_{1}^{-},t_{f}]$ and $c_{3}=x_{2d}(t_{c})$. For a steady state system, the Hamiltonian will satisfies
\begin{align}
 H=\lambda+c_{1}x_{2}(t)=0.
\label{3-38}
\end{align}
Refer to \eqref{2-12} and \eqref{3-38}, costate $p^{\rm T}(t)=[p_{1}(t)\quad p_{2}(t)]^{\rm T}$ will satisfy
\begin{equation}
\left\{
\begin{array}{l}
p_{1}(t)=-\frac{\lambda }{x_{2}(t)},\\ 
p_{2}(t)=\frac{\lambda }{x_{2}(t)}t+c_{2},
\label{3-39}
\end{array}
\right.
\end{equation}
According to Remark 1, $p_{2}(t_{1}^{-})$ will satisfy
\begin{align}
p_{2}(t_{1}^{-})=\frac{\lambda }{x_{2}(t_{1}^{-})}t_{1}^{-}+c_{2}=-1.
\end{align}
Note that, according to Remark 2,  $x_{2}(t_{1}^{-})=x_{2d}(t_{c})> 0$, so we can get the value of $c_{2}$:
\begin{align}
c_{2}=-1-\frac{\lambda }{x_{2d}}t_{1}^{-}< -1.
\end{align}
Furthermore, when $t\in [t_{1}^{-},\frac{x_{1d}(t_{1}^{-})-x_{1}(t_{1}^{-})}{x_{2d}(t_{c})}+t_{1}^{-}]$, $p_{2}(t)\leq 1$ always holds. And the equality $p_{2}(t)= 1$ holds if $t= \frac{x_{1d}(t_{1}^{-})-x_{1}(t_{1}^{-})}{x_{2d}(t_{c})}+t_{1}^{-}$. It means that the following equation needs to be satisfied:
\begin{align}
\frac{\lambda }{x_{2d}(t_{c})}t_{1}^{-}+\frac{\lambda \left (x_{1d}(t_{1}^{-})-x_{1}(t_{1}^{-}) \right )}{x_{2d}^{2}(t_{c})}-1-\frac{\lambda }{x_{2d}(t_{c})}t_{1}^{-}\leq 1.
\end{align}
Evidently,
\begin{align}
x_{1}(t_{1}^{-})\geq x_{1d}(t_{1}^{-})-\frac{2x_{2d}^{2}(t_{c})}{\lambda },
\end{align}
which completes the proof.\hfill $\Box$

\textit{Theorem 4:} If the control input sequence $\left \{-M,0\right \}$ is applied to plant \eqref{94-2}, there will be a switching instant $t_{1}^{+}$ in control process and the following conditions will be satisfied:
\begin{equation}
\left\{
\begin{array}{l}
x_{1}(t_{1}^{+})\leq x_{1d}(t_{1}^{+})+\frac{2x_{2d}^{2}(t_{c})}{\lambda },\\ 
x_{2}(t_{1}^{+})=x_{2d}(t_{c}).
\end{array}
\right.
\label{93-4}
\end{equation}

\textit{Proof:} Referring to Lemma 1, it is obvious that the control input sequence $u(t)=\left \{-M,0\right \}$ means $x_{2d}(t_{c})<0$. Similar to the proof of Theorem 3, we have
\begin{align}
u(t)=\left\{\begin{matrix}\vspace{0.2cm}
-M, & t\in [0,t_{1}^{+}],\\ \vspace{0.2cm}
 0, & t\in [t_{1}^{+},\frac{x_{1d}(t_{1}^{+})-x_{1}(t_{1}^{+})}{x_{2d}(t_{c})}+t_{1}^{-}].
\end{matrix}\right.
\end{align}
According to \eqref{3-39} and Remark 1, $p_{2}(t_{1}^{+})$ will satisfy
\begin{align}
p_{2}(t_{1}^{+})=\frac{\lambda }{x_{2}(t_{1}^{+})}t_{1}^{+}+c_{2}=1.
\end{align}
Note that, according to Remark 2, $x_{2}(t_{1}^{+})=x_{2d}(t_{c})< 0$, so we can get the value of $c_{2}$:
\begin{align}
c_{2}=1-\frac{\lambda }{x_{2d}}t_{1}^{+}>1.
\end{align}
Furthermore, when $ t\in [t_{1}^{+},\frac{x_{1d}(t_{1}^{+})-x_{1}(t_{1}^{+})}{x_{2d}(t_{c})}+t_{1}^{+}]$, $p_{2}(t)\geq- 1$ always holds. And the equality $p_{2}(t)=- 1$ holds if $t= \frac{x_{1d}(t_{1}^{+})-x_{1}(t_{1}^{+})}{x_{2d}(t_{c})}+t_{1}^{+}$. It means that the following equation needs to be satisfied:
\begin{align}
\frac{\lambda }{x_{2d}(t_{c})}t_{1}^{+}+\frac{\lambda \left (x_{1d}(t_{1}^{+})-x_{1}(t_{1}^{+}) \right )}{x_{2d}^{2}(t_{c})}+1-\frac{\lambda }{x_{2d}(t_{c})}t_{1}^{+}\geq-1.
\end{align}
Evidently,
\begin{align}
x_{1}(t_{1}^{+})\leq x_{1d}(t_{1}^{+})+\frac{2x_{2d}^{2}(t_{c})}{\lambda },
\end{align}
which completes the proof.\hfill $\Box$

On the basis of Theorem 1 and 3, we can find that there are two situations in which $u$ switches from $+M$ to 0. When the control input sequence is $\left \{+M,0,-M\right \}$, the system state at the switching instant $t_{1}^{-}$ satisfies the formula \eqref{3-24} and \eqref{93-1}, thus the switch curve is $Q_{3}$. When the control input sequence is $\left \{+M,0\right \}$, the switching time $t_{1}^{-}$ satisfies the formula \eqref{93-3}, thus the switch curve is $Q_{5}$ which intersects $Q_{3}$ at the point $\left (x_{1d}(t_{1}^{-})-\frac{2x_{2d}^{2}(t_{c})}{\lambda },x_{2d}(t_{c})\right )$ and intersects $Q_{1}$ at the point of target state. On the basis of Theorem 2 and 4, we can also find that there are two situations in which $u$ switches from $-M$ to 0 and the switch curve will be $Q_{4}$ and $Q_{6}$. In summary, the region $\Omega_{5}$ where $u=0$ can be expressed as the form in \eqref{93-18}. Note that increasing the value of $\delta$ will move the switching curves $Q_{5}$ and $Q_{6}$, so that the system can reach the target faster.

Another difference between QTFOC and TFOC for static targets is that there exist a new plane area which can be named as transient loop. It is equivalent to $\left \{Q_{1}<0\cap Q_{2}>0\right \}$. The system cannot reach the target state in the transient loop because $\left | x_{2}(t)\right |<\left | x_{2d}(t_{c})\right |$ so that we can choose $x_{1}$ as the first control target and $x_{2}$ as the auxiliary control target, hence the control input in transient loop can be chosen as $+M$ if $Q_{8}<0$ and $-M$ if $Q_{8}>0$. Combining the analysis on $\Omega_{5}$, we can naturally obtain $\Omega_{1}$ and $\Omega_{3}$.

However, although the local linear controller \eqref{93-19} can eliminate the highly oscillatory behavior of the system near the target, it can also be triggered by the feedback delay of the target state near the switch curves. In other words, the system will oscillate due to the changes of control input near the switch curves $Q_{1}$ and $Q_{2}$ when $x_{2d}(t_{c})>0$ and $x_{2d}(t_{c})<0$ respectively. Therefore, we designed a buffer area near $Q_{1}$, the width of which is $\xi$. When the system state leaves $Q_{1}$ to the right plane, the value of $u(t)$ gradually increases, and when it exceeds this buffer, the value of $u(t)$ becomes $M$. This is the region $\Omega_{2}$ in \eqref{93-15} and in the same way we can obtain $\Omega_{4}$ in \eqref{93-17}.

So far, all the regions in \mbox{Fig. \ref{FIG_5}} have been analyzed. Next, we will consider the actual servo system in this brief which is named as visual tracking system. and Section IV will focus on the impact of Coulomb friction on the control strategy.

\section{The Influence of Coulomb Friction Load Asymmetry on Visual Tracking System}
This section mainly studies how to maximize the limit performance of the system under the condition of asymmetric load. The following model is not only applicable to the visual tracking system driven by a servo motor, but also to a variety of actuators.
\begin{figure}[!t]
\centerline{\includegraphics[width=0.8\columnwidth]{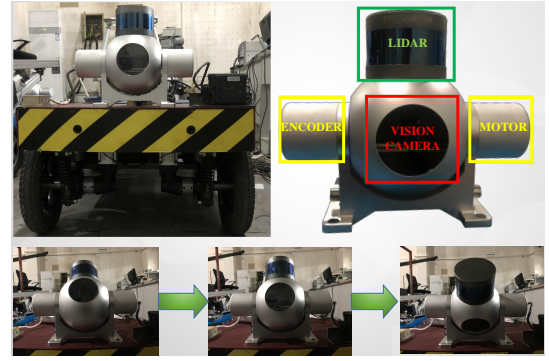}}
\caption{Visual tracking system mounted on the unmanned vehicle.}
\label{FIG_4}
\end{figure}
\begin{align}
J\alpha =J\ddot{\theta }=T_{d}-B\dot{\theta }-K\theta-T_{f},
\label{5-65}
\end{align}
where $J$ denotes the moment of inertia of the visual tracking system. $\theta $ and $\alpha $ represent the rotation angle and rotation angular acceleration, respectively. $B$ is the damping coefficient (including mechanical damping, electromagnetic damping, etc.). $K$ denotes the elastic coefficient of the system. $T_{f}$ indicates the Coulomb friction moment, and $T_{d}$ indicates the electromagnetic torque generated by the servo motor.

We focuses on a visual tracking system (see \mbox{Fig. \ref{FIG_4}}) for tracking dynamic targets, which has a small rotation angle range (i.e., $\left [-120^{\circ},+120^{\circ}\right ]$) and a low tracking speed, but requires a high angular acceleration and often works at the maximum output torque. Therefore, compared with the Coulomb friction load and inertial load, the damping load proportional to the rotation speed and the elastic load proportional to the rotation angle are small enough to be ignored in the following analysis.
Note that the Coulomb friction loads present asymmetric characteristics relative to the motion process, which will adversely affect the acceleration performance of the system but benefit the deceleration performance.

On the basis of the above analysis, we can transform $\eqref{5-65}$ into a form similar to a double integrator, the model of visual servo system is rewritten as $\dot{x_{1}}=x_{2}$, $\dot{x_{2}}=\alpha $, where ${{x}_{1}}$ and ${{x}_{2}}$ are equivalent to $\theta $ and $\dot{\theta }$, respectively. ${{T}_{d}}$ ranges in $[-JM-{{T}_{f}},JM+{{T}_{f}}]$. Therefore, $\alpha \in [-M,M]$ when the system is accelerating and $\alpha \in [-M-{{T}_{f}},-M+{{T}_{f}}]$ when the system is slowing down. That is, when the system is decelerating and deviates from the switch curves ${{Q}_{1}}$ or ${{Q}_{2}}$, ${{T}_{f}}$ contributes to deceleration. In order to be consistent with the previous analysis, we rewrite $\alpha $ to $u(t)$ here. When $\left \{Q_{2}+\xi\leq 0\cap x_{2}<0\right\}$, the allowable maximum value of $u(t)$ can be modified to $u_{max}(t)=M+K_{1}$. When $\left \{Q_{1}-\xi\geq  0\cap x_{2}>0\right\}$, the allowable minimum value of $u(t)$ can be modified to $u_{min}(t)=-M-K_{1}$. The value of $K_{1}$ depends on the asymmetric $T_{f}$ in \eqref{5-65}. This is a engineering method, but it often plays a satisfactory role in the actual application process.

In addition, this method only works when the system is decelerating, so a buffer area with width $\frac{K_{1}}{M}\xi$ similar to that in Section IV is set up.

According to the above discussion, the value of $u(t)$ proposed in \eqref{93-15} and \eqref{93-17} is redefined as follows:

\begin{enumerate}
\item  
$u(t)= {M}'min\left \{1,\left | \frac{Q_{2}}{{\xi}'}\right |\right \}$,  iff $(x_{1},x_{2} )\in\Omega_{2}$,
\item 
$u(t)=-{M}'min\left \{1,\left | \frac{Q_{1}}{{\xi}'}\right |\right \}$,  iff $(x_{1},x_{2} )\in\Omega_{4}$,

\end{enumerate}
where ${M}'=M+K_{1}$ and ${\xi}'=\frac{M+K_{1}}{M}\xi$.

\section{Experimental Validation}

In this section, the performance of the proposed controller (QTFOC) is evaluated by using an experimental setup (see \mbox{Fig. \ref{FIG_4}}), which consists of lidar, vision sensor, and execution motor, to show the efficacy and superiority of the control strategy and verify the theoretical results. Considering the vertical scanning field of the lidar is limited, when the lidar and the vision camera perform fusion detection, the lidar and the vision camera can obtain high-altitude target information through the rotation of the turntable. Therefore, the turntable needs to perform large-angle adjustments based on the given information to aim at targets beyond the vertical scanning field of the lidar, and then track the movement of targets in a small range. 

On the basis of Section IV, the model of the visual tracking system is shown below:
\begin{align}
\left\{\begin{matrix}
\dot{\theta }(t)=\omega(t) \\ 
\dot{\omega }(t)=\frac{1}{J}(T_{d}(t)-T_{f}).
\end{matrix}\right.
\end{align}

\begin{table}
\caption{Experimental Setup Parameters}
\label{table_1}
\setlength{\tabcolsep}{3pt}
\begin{tabular}{|p{25pt}|p{85pt}|p{115pt}|}
\hline
Symbol& 
Quantity& 
Value \\
\hline
$ J $& 
moment of inertia& 
$ 0.0186(kgm^{2}) $  \\
$ T_{d} $& 
electromagnetic torque&
$ \left [-0.125,0.125 \right ](Nm) $  \\
$ T_{f} $& 
 friction resistance moment & 
$ \pm 0.0084(Nm) $\\
$ \xi $& 
buffer width& 
$ 0.4 $ \\
$ \delta  $& 
offset& 
$ 1 $ \\

\hline

\end{tabular}
\label{tab1}
\end{table}

The relevant parameters of the experimental setup are presented in \mbox{Table \ref{table_1}}. From the Section IV and \mbox{Table \ref{table_1}}, we can know that $\dot{\omega } \in [-359.2,359.2]$ when the system is accelerating and $\dot{\omega } \in [-410.9,410.9]$ when the system is slowing down.

 In the aspect of feedback communication, all the necessary data are sent to controller every 3ms by the serial communication port of the STM32 with the baud rate being 115200. Note that, for actual systems, the signals of sensors are often accompanied by noise and sometimes there are no available sensors to measure output derivatives of a system. Therefore, extended kalman filter \cite{ljung1979asymptotic, lu2016improved, kim2018introduction} and tracking differentiator \cite{han2009pid, levant1998robust, davila2013exact} may be required. This is only a signal processing mechanism independent of the control strategy, so it will not be described here. Readers interested in this aspect can refer to the above papers. It is noteworthy that for the STM32 microcontroller, the computational complexity must be considered. Unlike classical process control applications, the sampling time of the fast dynamic systems (embedded systems) is typically of millisecond order. The research results in \cite{2019Fast} showed that the classical Dynamic Matrix Control (DMC) and Generalized Predictive Control (GPC) algorithms is unsuitable for STM32, since calculations last longer than the sampling period. The proposed QTFOC can obtain an explicit solution, and the calculation time required for this algorithm is very short. Therefore, this is beneficial to reduce the computational burden of STM32. In addition, the system shown in \mbox{Fig. \ref{FIG_4}} is powered by lithium batteries. Significantly, the capacity of a lithium battery refers to the limited amount of charge released when the battery is fully discharged, that is, the amount of charge that can be released is fixed when a battery is fully charged. Since the intensity of the current is the quantity of charge which passes in a conductor per unit of time, for an servo motor powered by a lithium battery, the integral of $\left | u\right |$ with respect to time, which is proportional to the current intensity, can reflect the power loss of the lithium battery during the control process. Similar researches on the systems powered by lithium batteries have been introduced in \cite{erdinc2009dynamic} and \cite{rong2006analytical}.

After completing the above preparations, we conduct following experiments. We will let the experimental setup switch and track between three dynamic targets, which are named "Object A", "Object B", and "Object C". The trajectories of "Object A", "Object B", and "Object C" are $5sin(\pi t)+10sin(2t)$, $50+5sin(\pi t)+10sin(2t)$, and $-50+5sin(\pi t)+10sin(2t)$ respectively. For the convenience of description, we establish an experimental scene named "CASE-F". It is described as following: In 0 to 5 seconds, the visual tracking system is required to track "Object A". In 5-10 seconds, the visual tracking system is required to track "Object B". After 10 seconds, the visual tracking system is required to track "Object C". Therefore, "CASE-F" can evaluate the speed of the system switching between multiple dynamic targets and the dynamic tracking performance of the system under the corresponding control strategy.

In order to better demonstrate the effect of the control strategy, we compare the performance of QTFOC for dynamic targets with TFOC and proportional-integral-derivative (PID) control law. The PID control law is improved by compensator, which can resist the effects of differential noise. The transfer function is as follows:
\begin{align}
G(s)=P+I\frac{1}{s}+D\frac{N}{1+N\frac{1}{s}}.
\end{align}
In addition, the control input of the PID controller should satisfies $\left | u\right |\leq M$, where $M=359.2$. After actual experimental testing, two sets of representative PID parameters are selected:
\begin{align}
&PID1:\: K_{p1}=45.0, \: K_{i1}=3.51, \: K_{d1}=9.45,\: N_{1}=10,\nonumber\\
&PID2:\: K_{p2}=24.3, \: K_{i2}=2.04, \: K_{d2}=11.51, N_{2}=10.\nonumber
\end{align}

\begin{figure}[!t]
\centerline{\includegraphics[width=\columnwidth]{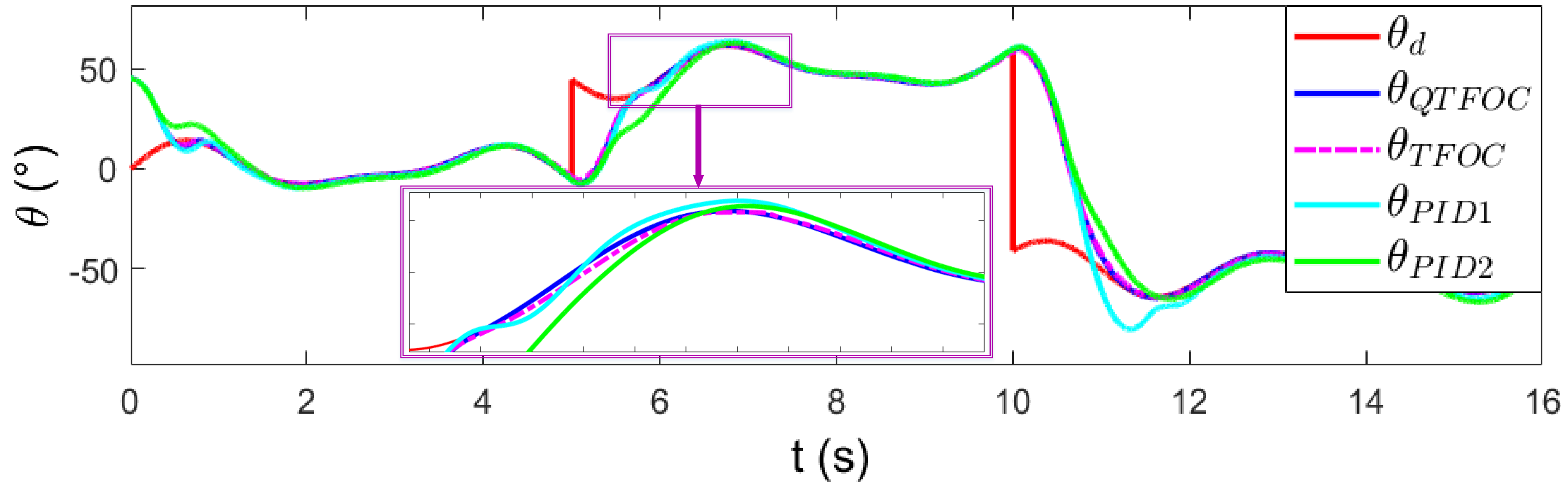}}
\caption{Time histories of reference angle and output angles under different control strategies.}
\label{FIG_7-11}
\end{figure}

\begin{figure}[!t]
\centerline{\includegraphics[width=\columnwidth]{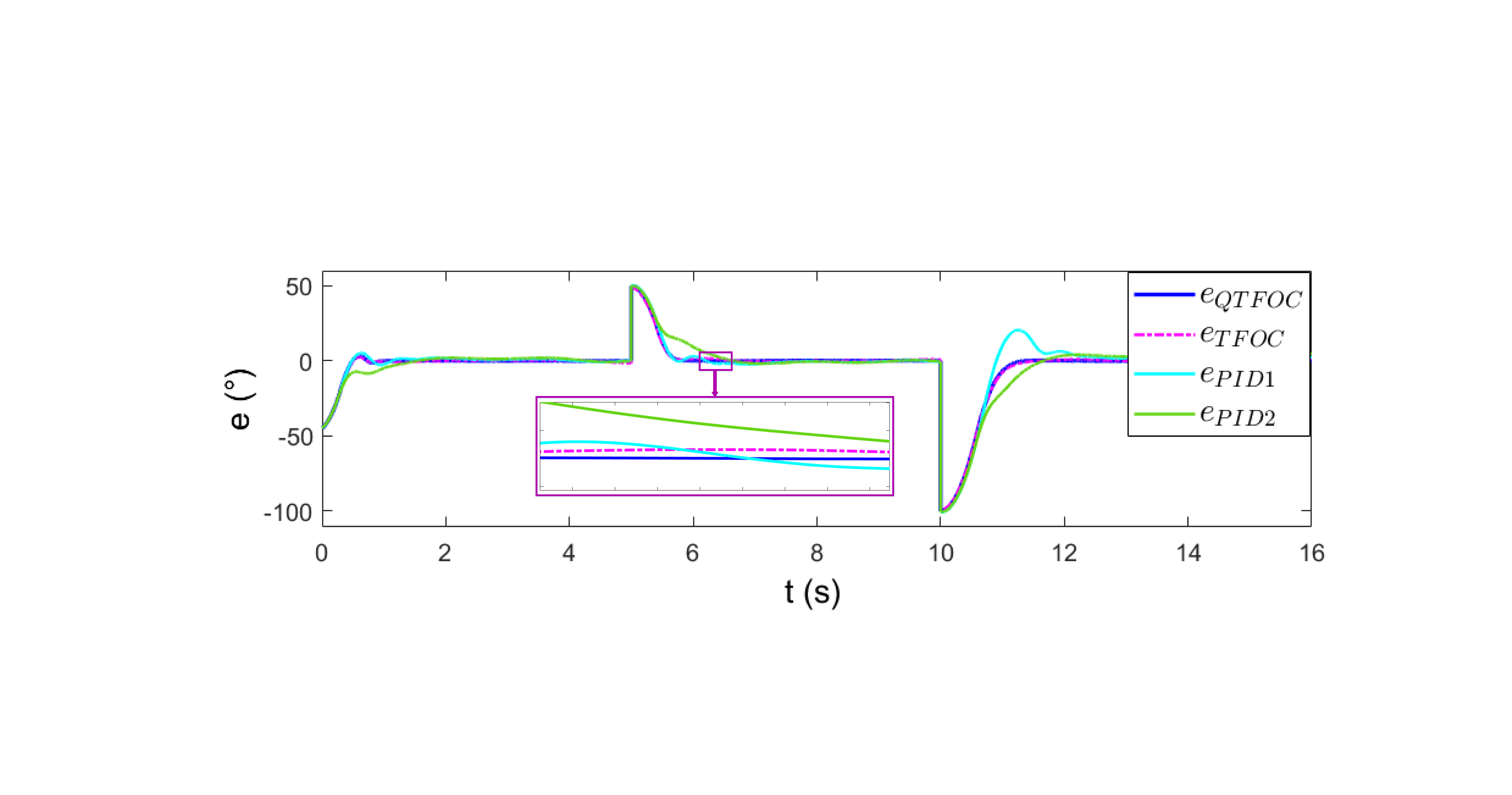}}
\caption{Time histories of tracking errors under different control strategies. }
\label{FIG_7-12}
\end{figure}

\begin{figure}[!t]
\centerline{\includegraphics[ width=\columnwidth]{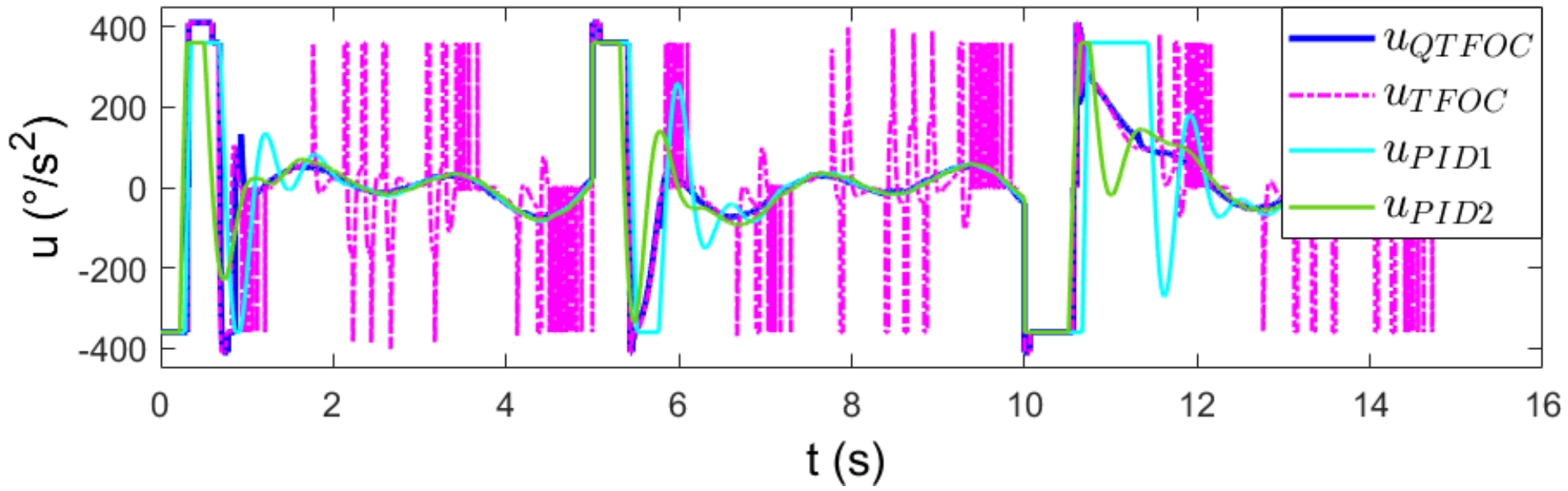}}
\caption{Time histories of control inputs under different control strategies.}
\label{FIG_7-13}
\end{figure}

\begin{figure}[!t]
\centerline{\includegraphics[ width=\columnwidth]{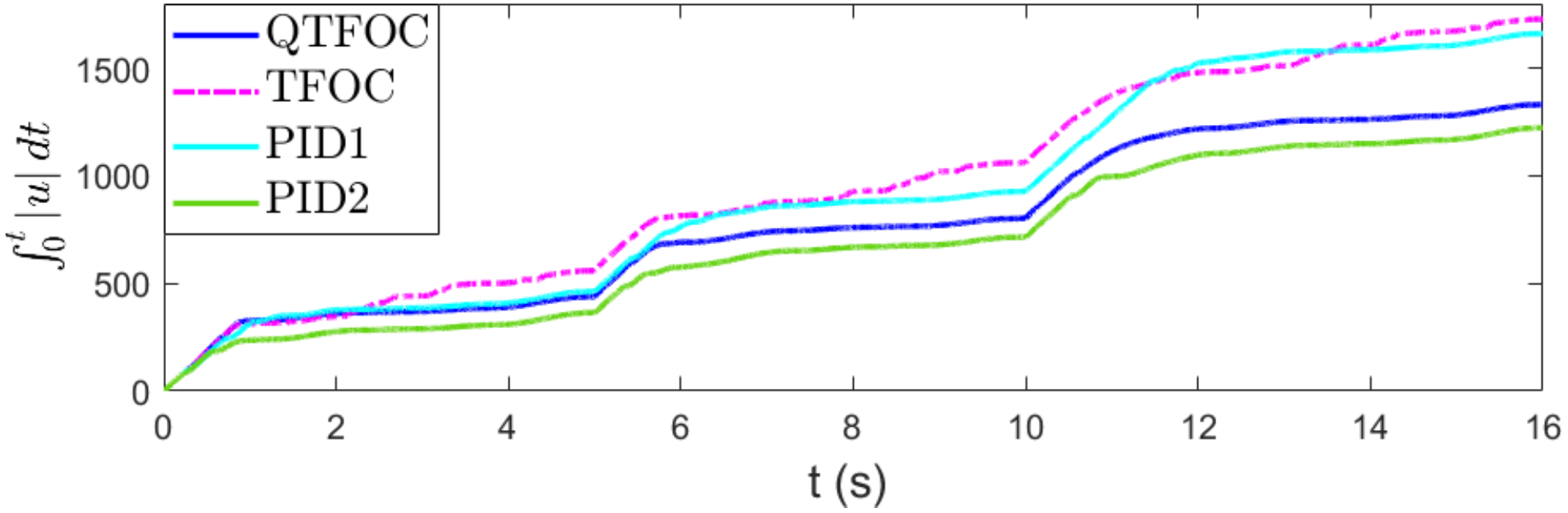}}
\caption{The integrals of $\left |u \right |$ with respect to $t$ under different control strategies.}
\label{FIG_7-14}
\end{figure}

The initial states of visual tracking system are given as $\theta (0)=45^{\circ}$, $\omega(0)=0^{\circ}/s$ and the experimental setup is required to track "CASE-F". The time weight $\lambda$ is set to $\lambda = 10000$. The experimental performance of QTFOC, TFOC, PID1, and PID2 are shown in \mbox{Fig. \ref{FIG_7-11}-\ref{FIG_7-14}}, which demonstrate the superiority of QTFOC in dynamic target tracking. 
Moreover, as shown in \mbox{Fig. \ref{FIG_7-13}}, the control input of TFOC will switch frequently because of the speed mismatch and feedback delays.
\mbox{Fig. \ref{FIG_7-14}} verifies that the frequent switching of the TFOC control input (see \mbox{Fig. \ref{FIG_7-13}}) results in a larger accumulation of $\left| u \right|$, that is, TFOC will cause the system’s lithium battery to consume more power with the same $\lambda $ value.
In summary, TFOC is inapplicable when tracking dynamic targets, and QTFOC is feasible for this. 
On the other hand, compared with QTFOC, when two sets of PID controllers with different parameters are used to track "CASE-F", the switching and tracking performance will be worse due to input saturation constraints which are illustrated in \mbox{Fig. \ref{FIG_7-13}}. In addition, from \mbox{Fig. \ref{FIG_7-11}} and \mbox{Fig. \ref{FIG_7-12}}, we can see that in the process of switching the target, QTFOC can recover the tracking of the target faster than PID1 and PID2.

\begin{figure}[!t]
\centerline{\includegraphics[ width=\columnwidth]{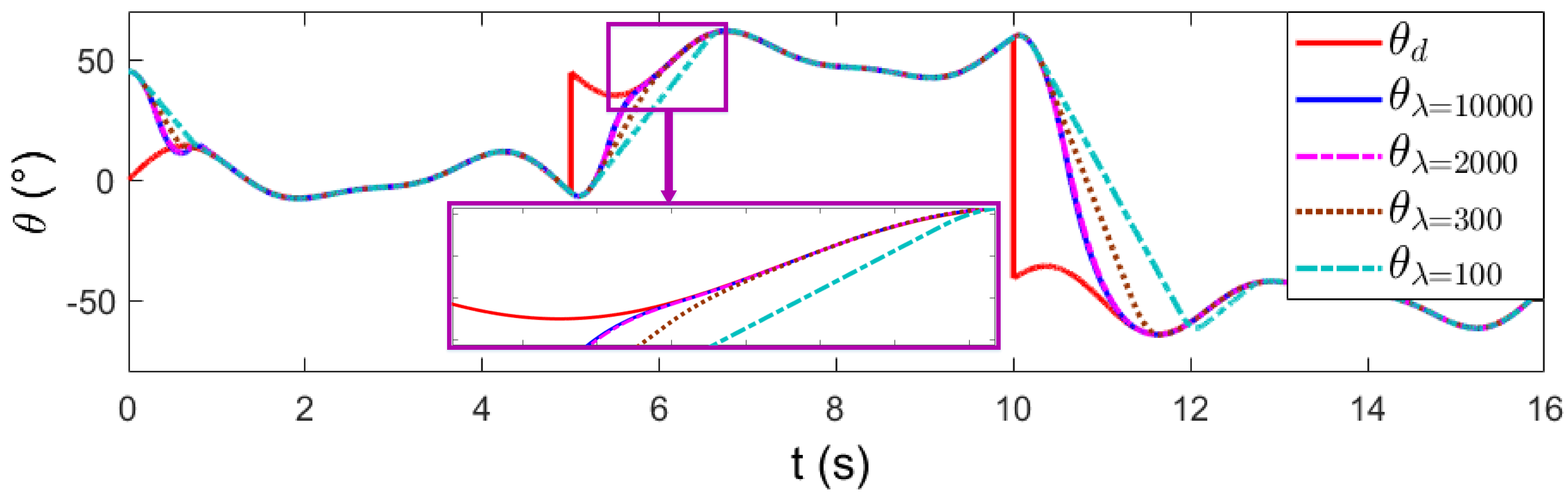}}
\caption{Time histories of reference angle and the QTFOC's output angles with different $\lambda$.}
\label{FIG_7-19}
\end{figure}

\begin{figure}[!t]
\centerline{\includegraphics[ width=\columnwidth]{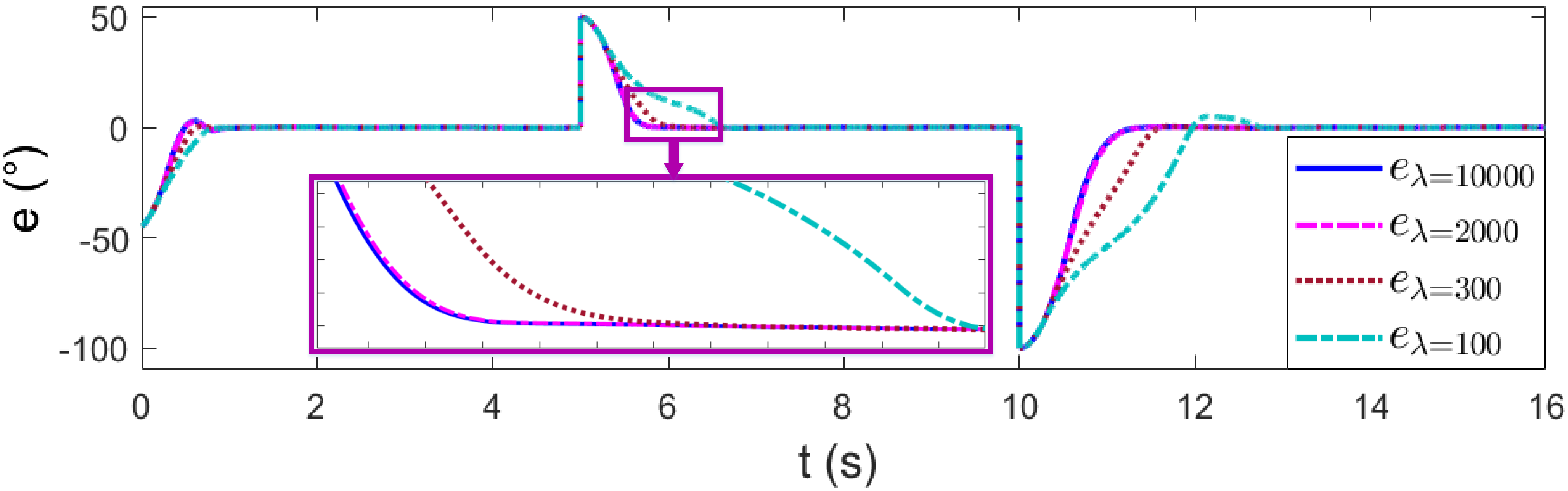}}
\caption{Time histories of the QTFOC's tracking errors with different $\lambda$.}
\label{FIG_7-20}
\end{figure}

\begin{figure}[!t]
\centerline{\includegraphics[ width=\columnwidth]{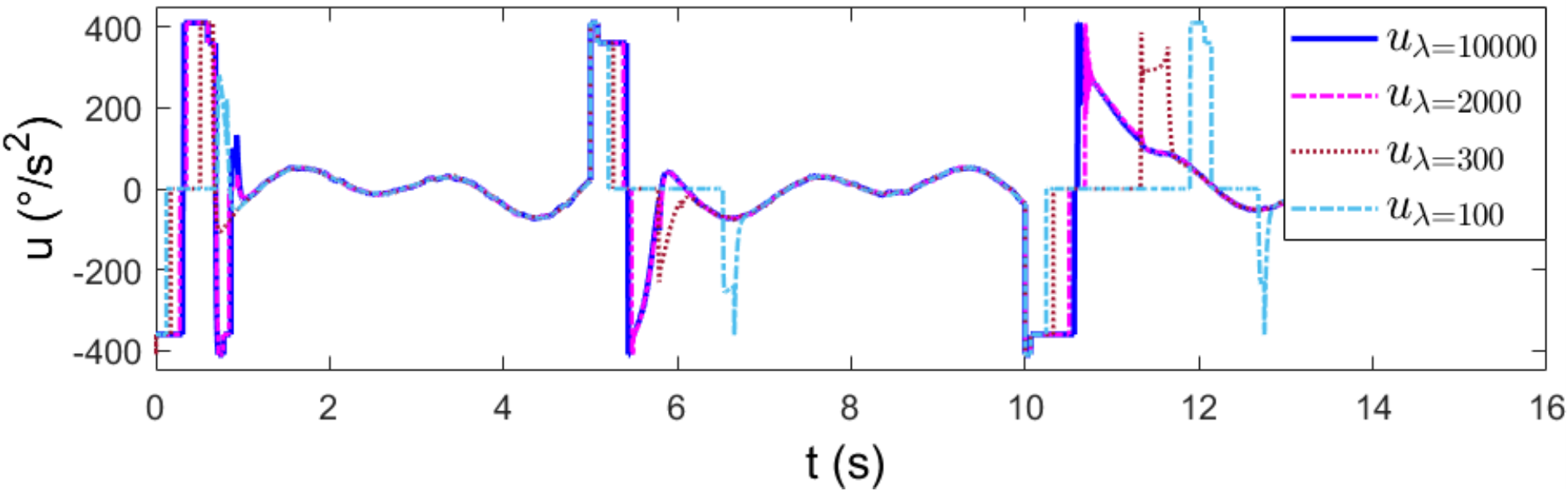}}
\caption{Time histories of the QTFOC's control inputs with different $\lambda$.}
\label{FIG_7-21}
\end{figure}

\begin{figure}[!t]
\centerline{\includegraphics[ width=\columnwidth]{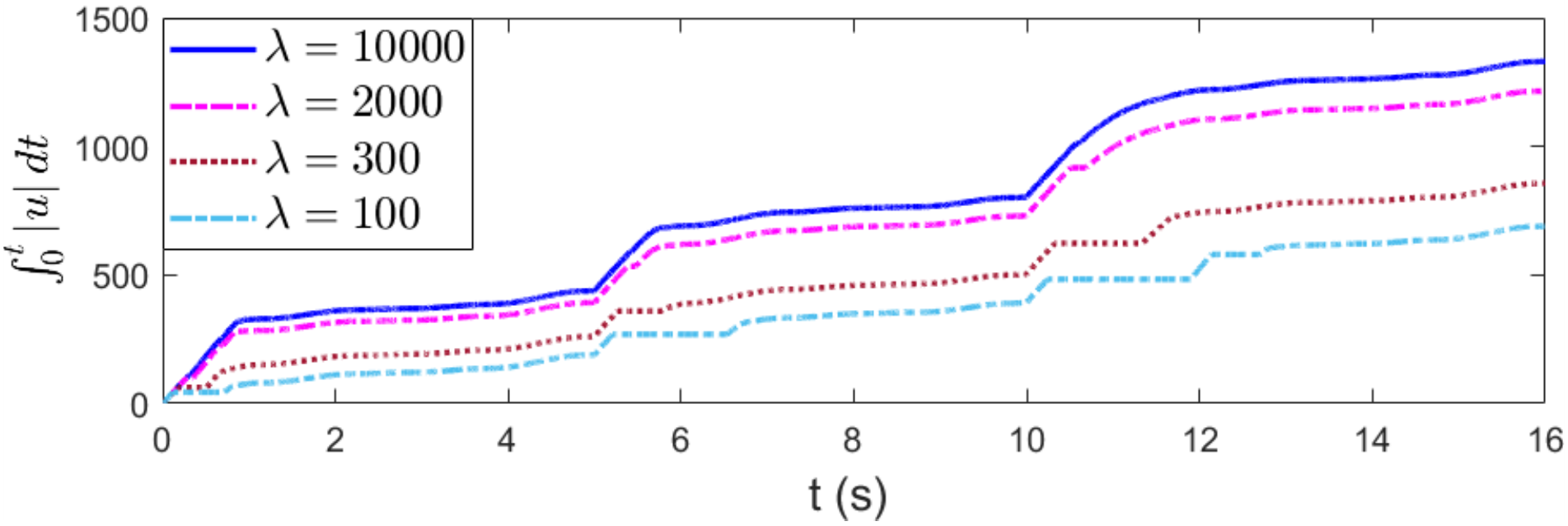}}
\caption{The integrals of $\left |u \right |$ with respect to $t$ with different $\lambda$.}
\label{FIG_7-22}
\end{figure}

The influence of different $\lambda$ on system tracking performance and fuel consumption are shown in \mbox{Fig. \ref{FIG_7-19}-\ref{FIG_7-22}}. The initial states are given as $\theta (0)=45^{\circ}$, $\omega(0)=0^{\circ}/s$, and the experimental setup was also required to track "CASE-F". It can be seen from \mbox{Fig. \ref{FIG_7-19}} and \mbox{Fig. \ref{FIG_7-20}} that increasing the value of $\lambda$ will increase the response speed of the system, which is beneficial to quickly reduce the tracking error when switching targets. Combining the above results in \mbox{Fig. \ref{FIG_7-21}} and \mbox{Fig. \ref{FIG_7-22}}, we can conclude that when increasing the value of $\lambda$ to obtain a faster response speed, QTFOC will make the the system’s lithium battery consume more power, which will reduce the endurance of the unmanned system to a certain extent.
In other words, the maximum continuous working time of the unmanned system can be increased by reducing the value of  $\lambda$.

\section{Conclusion}
In this work, a dynamic target tracking problem is studied in the framework of time-fuel optimal control strategy. The experiments are implemented using a visual tracking system, and the results shown in \mbox{Fig. \ref{FIG_7-11}-\ref{FIG_7-20}} have demonstrated the superiority of our proposed QTFOC in dealing with dynamic target tracking problems, especially the multi-target switching tracking problems. What's more, QTFOC can also adjust the weight of response speed, which making the unmanned system perform better under different working conditions. A further step is to investigate how to use the funnel control strategy to improve the local linear control strategy, which is a non-trivial extension of the current work. Future research work also includes robustness analysis of the proposed algorithm and dynamic target trajectory prediction among others.

\appendices
\section{Stability Analysis of the Quasi Time-Fuel Optimal Control Strategy}
The analysis consists of two parts:
\begin{enumerate}
	\item First, we show that all trajectories originating outside the local linear region ($\Omega_{6}$) will converge to the target point, i.e., will enter the local linear region.  
	\item Second, any trajectory in the local linear region ($\Omega_{6}$) will remain there indefinitely and once in $\Omega_{6}$, the trajectory converges to the target point.
\end{enumerate}

\textbf{Part One:} In the beginning we set $K_{1}=0$, $ \xi =0$, and select the following Lyapunov function:
	\begin{align}
	    V=\left\{\begin{matrix} 
f_{1}(x_{1e},x_{2e}), & x_{1e}<-\sigma_{x},\\ 
 g_{1}(x_{1e},x_{2e}),& \left |x_{1e} \right |\leq \sigma_{x} ,\\ 
 f_{2}(x_{1e},x_{2e}),& x_{1e}>\sigma_{x},
\end{matrix}\right.
	\end{align}
where $\sigma _{x}>0$, $x_{1e}(t)=x_{1}(t)-x_{1d}(t)$, $x_{2e}(t)=x_{2}(t)-x_{2d}(t_{c})$, and
    \begin{align}
f_{1}(x_{1e},x_{2e})&=[x_{1e}-\frac{1}{2M}x_{2e}^{2}]^{2}+1,\\
f_{2}(x_{1e},x_{2e})&=[x_{1e}+\frac{1}{2M}x_{2e}^{2}]^{2}+1,\\
g_{1}(x_{1e},x_{2e})&=g_{1+}(x_{1e},x_{2e})\quad x_{2e}>0,\\
g_{1}(x_{1e},x_{2e})&=g_{1-}(x_{1e},x_{2e})\quad x_{2e}<0. 
\end{align}

Meanwhile, ${{g}_{1}}({{x}_{1e}},{{x}_{2e}})>0$ when $\left| {{x}_{1e}} \right|\le {{\sigma }_{x}}.$ Note that when $x_{1e}<-\sigma _{x}$,
\begin{align}\dot{V}=\dot{f}_{1} (x_{1e},x_{2e})=[\frac{2(M-u)}{M}x_{1e}+\frac{u-M}{M^{2}}x_{2e}^{2}]x_{2e}
\label{93-52}
\end{align}
Then on the basis of the control strategy \eqref{93-6}-\eqref{93-18} in this paper, we can obtain that $u\le M$ when $x_{1e}<-\sigma _{x}$ and ${{x}_{2e}}>0$; and hence, $\dot{V}<0$ in this region. However, $u(t)\not\equiv M$ in this case, which means $\dot{V}\not\equiv 0$. Similarly, $u=M$ when ${{x}_{1e}}<-{{\sigma }_{x}}$ and ${{x}_{2e}}<0$; and hence, $\dot{V}=0$ in this region. 

When ${{x}_{1e}}>{{\sigma }_{x}}$,
\begin{align}\dot{V}=\dot{f}_{2} (x_{1e},x_{2e})=[\frac{2(M+u)}{M}x_{1e}+\frac{u+M}{M^{2}}x_{2e}^{2}]x_{2e}.
\label{93-53}
\end{align}
Then on the basis of the control strategy \eqref{93-6}-\eqref{93-18} in this paper, we can obtain that $u=-M$ when ${{x}_{1e}}>{{\sigma }_{x}}$ and ${{x}_{2e}}>0$; and hence, $\dot{V}=0$ in this region. Similarly, $u\geq -M$ when ${{x}_{1e}}>{{\sigma }_{x}}$ and ${{x}_{2e}}<0$; and hence, $\dot{V}<0$ in this region. Moreover, $u(t)\not\equiv-M$ in this case, which means $\dot{V}\not\equiv0$.

On the other hand, we note that
\begin{align}
    \lim_{\sigma _{x}\rightarrow 0}f_{1}(-\sigma _{x},x_{2e})=\lim_{\sigma _{x}\rightarrow 0}f_{2}(\sigma _{x},x_{2e}).
\end{align}
When ${{x}_{2e}}>0$ and $\left |{{x}_{1e}} \right |\leq \sigma _{x}$, substituting the value of control input $u$ into \eqref{93-52} and \eqref{93-53} can obtain that ${{\dot{f}}_{1}}({x}_{1e},{{x}_{2e}})\leq 0$, ${{\dot{f}}_{1}}({x}_{1e},{{x}_{2e}})\not\equiv 0$ and ${{\dot{f}}_{2}}({{x}_{1e}},{{x}_{2e}})=0$. Thus, we assume that there exists a differentiable function ${{g}_{1+}}({{x}_{1e}},{{x}_{2e}})$ such that the following conditions hold:
\begin{align}
    &{{g}_{1+}}(-{{\sigma }_{x}},{{x}_{2e}})={{f}_{1}}(-{{\sigma }_{x}},{{x}_{2e}})>0,
    \\&{{g}_{1+}}({{\sigma }_{x}},{{x}_{2e}})={{f}_{2}}({{\sigma }_{x}},{{x}_{2e}})>0,
    \\ &{{g}_{1+}}({{x}_{1e}},{{x}_{2e}})>0,
\end{align}
and the derivative of ${{g}_{1+}}({{x}_{1e}},{{x}_{2e}})$ satisfies:
\begin{align}
    &{{\dot{g}}_{1+}}(-{{\sigma }_{x}},{{x}_{2e}})=\dot{f}_{1}(-{\sigma }_{x},{x}_{2e}),
    \\&{{\dot{g}}_{1+}}({{\sigma }_{x}},{{x}_{2e}})=\dot{f}_{2}({\sigma }_{x},{x}_{2e}),
    \\&{{\dot{g}}_{1+}}({{x}_{1e}},{{x}_{2e}})\le 0.
\end{align}
Moreover, when ${{x}_{2e}}<0$and $\left |{{x}_{1e}} \right |\leq \sigma _{x}$, substituting the value of control input $u$ into \eqref{93-52} and \eqref{93-53} can obtain that ${{\dot{f}}_{1}}({{x}_{1e}},{{x}_{2e}})=0$, ${{\dot{f}}_{2}}({{x}_{1e}},{{x}_{2e}})\leq0$ and ${{\dot{f}}_{2}}({{x}_{1e}},{{x}_{2e}})\not\equiv 0.$ Thus, we assume that there exists a differentiable function ${{g}_{1-}}({{x}_{1e}},{{x}_{2e}})$ such that the following conditions hold:
\begin{align}
    &{{g}_{1-}}(-{{\sigma }_{x}},{{x}_{2e}})={{f}_{1}}(-{{\sigma }_{x}},{{x}_{2e}})>0,
    \\&{{g}_{1-}}({{\sigma }_{x}},{{x}_{2e}})={{f}_{2}}({{\sigma }_{x}},{{x}_{2e}})>0,
    \\ &{{g}_{1-}}({{x}_{1e}},{{x}_{2e}})>0,
\end{align}
and the derivative of ${{g}_{1-}}({{x}_{1e}},{{x}_{2e}})$ satisfies:
\begin{align}
    &{{\dot{g}}_{1-}}(-{{\sigma }_{x}},{{x}_{2e}})=\dot{f}_{1}(-{\sigma }_{x},{x}_{2e}),
    \\&{{\dot{g}}_{1-}}({{\sigma }_{x}},{{x}_{2e}})=\dot{f}_{2}({\sigma }_{x},{x}_{2e}),
    \\&{{\dot{g}}_{1-}}({{x}_{2e}},{{x}_{2e}})\leq 0.
\end{align}
Therefore, there exists a continuous and differentiable Lyapunov function $V$, where
\begin{align}
    V=\left\{\begin{matrix} 
f_{1}(x_{1e},x_{2e}), & x_{1e}<-\sigma_{x},\\ 
 g_{1}(x_{1e},x_{2e}),& \left |x_{1e} \right |\leq \sigma_{x} ,\\ 
 f_{2}(x_{1e},x_{2e}),& x_{1e}>\sigma_{x},
\end{matrix}\right.
\end{align}
such that $V>0$and $\dot{V}\leq 0$ hold, and for any nonzero point $\boldsymbol{x_{e}}$ in the state space, $\dot{V}(\boldsymbol{x_{e}})\not\equiv0$. Hence, the system is asymptotically stable and all trajectories originating outside the local linear region ($\Omega_{6}$) will converge to the target point, i.e., will enter the local linear region.

When $K_{1}$ and $\xi$ are not equal to 0, we can make the range ( $\left | x_{1e}\right |\leq \sigma _{x}$ ) of $g_{1}$ include the boundary of the buffer areas created by $\xi$ and then let
\begin{align}f_{1}(x_{1e},x_{2e})=[x_{1e}-\frac{1}{2(M+K_{1})}x_{2e}^{2}]^{2}+1,\\
f_{2}(x_{1e},x_{2e})=[x_{1e}+\frac{1}{2(M+K_{1})}x_{2e}^{2}]^{2}+1,
\end{align}
and the above conclusion can be proved by similar methods. However, for the piecewise control system in this paper, it is indeed difficult to find a completely analytical Lyapunov function but we will continue to strive towards this goal in the future work.

\textbf{Part two:} In the revised manuscript, the region and the control strategy of the local linear controller are set as follows:
Boundary of the region:
\begin{align}
    Q_{9}&=a_{1}(x_{1}(t)-x_{1d}(t))^{2} +a_{4}(x_{2}(t)-x_{2d}(t_{c}))^{2}
    \nonumber\\&+2a_{2}(x_{1}(t)-x_{1d}(t))(x_{2}(t)-x_{2d}(t_{c}))- k_{F},
    \label{93-70}
\end{align}
where
\begin{align}
k_{F}&=\frac{Ma_{4}(a_{1}-\sqrt{M}a_{2}+\frac{Ma_{4}}{4})-(\sqrt{M}a_{2}-\frac{Ma_{4}}{2})^{2}}{4(a_{1}-\sqrt{M}a_{2}+\frac{Ma_{4}}{4})}, \label{93-71}\\
a_{1}&=\frac{M+5}{4\sqrt{M}},\quad
a_{2}=\frac{1}{2M},\quad
a_{4}=\frac{1+M}{4M\sqrt{M}}.\label{93-72}
\end{align}
The $u(t)$ of the local linear controller:
\begin{align}
    u(t)=-M\left [(x_{1}(t)-x_{1d}(t))+\frac{2}{\sqrt{M}}(x_{2}(t)-x_{2d}(t_{c}))\right ].
\end{align}

In accordance with the form of local linear controller, the unsaturation of $u(t)$ can be described by the following formula:
\begin{align}
    -M\leq -M\left [ x_{1e}(t)+\frac{2}{\sqrt{M}}x_{2e}(t)\right ]\leq M.
\end{align}
This indicates that $\left |u(t) \right |\leq M$ can be satisfied in $\Omega_{6}$ ( $Q_{9}\leq 0$ ) as long as the solutions of the following two equations do not have two different real roots respectively, that is, the elliptic region does not intersect the unsaturated boundary of $u(t)$.
\begin{align}
    Q_{9}=\left [ x_{1e}(t)+\frac{2}{\sqrt{M}}x_{2e}(t)\right ], \label{93-75}\\
    Q_{9}=-\left [ x_{1e}(t)+\frac{2}{\sqrt{M}}x_{2e}(t)\right ].
    \label{93-76}
\end{align}
Therefore, in order to meet the above conditions, we substitute \eqref{93-70} into \eqref{93-75} and \eqref{93-76} to eliminate $x_{2e}(t)$, and the following result can be obtained:
\begin{align}
    \left ( a_{1}-\sqrt{M}a_{2}+\frac{Ma_{4}}{4}\right )x_{1e}^{2}(t)&+\left (\sqrt{M}a_{2}-\frac{Ma_{4}}{2} \right )x_{1e}(t)\nonumber\\&+\frac{Ma_{4}}{4}-k_{F}\leq 0,
\end{align}
that is 
\begin{align}
    k_{F}\leq \frac{Ma_{4}(a_{1}-\sqrt{M}a_{2}+\frac{Ma_{4}}{4})-(\sqrt{M}a_{2}-\frac{Ma_{4}}{2})^{2}}{4(a_{1}-\sqrt{M}a_{2}+\frac{Ma_{4}}{4})},
    \label{93-78}
\end{align}
Note that the parameter $K_{F}$ designed by the local linear controller in \eqref{93-71} satisfied the condition of \eqref{93-78}, so the control input never saturates for $x\in \Omega_{6}$.

On the basis of \eqref{93-70}, the region  of the local linear controller is rewritten as follows:    
\begin{align}
\boldsymbol{x_{e}}^{\rm T}\boldsymbol{P}\boldsymbol{x_{e}}\leq k_{F}, 
\end{align}
where
\begin{align}
\boldsymbol{x_{e}}=\begin{bmatrix}
x_{1e}\\ 
x_{2e}
\end{bmatrix},\qquad
\boldsymbol{P}=\begin{bmatrix}
a_{1} & a_{2}\\ 
 a_{2}& a_{3} 
\end{bmatrix},
\end{align}
and the parameters of $\boldsymbol{P}$ are described in \eqref{93-72}. Note that for all $M>0$, the matrix $\boldsymbol{P}$ is positive definite. Thus, a Lyapunov function that shows the stability of the system in $\Omega_{6}$ is
\begin{align}
V=\boldsymbol{x_{e}}^{\rm T}\boldsymbol{P}\boldsymbol{x_{e}}>0. 
\label{93-81}
\end{align}
Differentiating \eqref{93-81} gives
\begin{align}
\dot{V}=\boldsymbol{x_{e}}^{\rm T}(\boldsymbol{A_{cl}^{\rm T}P+PA_{cl}})\boldsymbol{x_{e}}, 
\end{align}
where
\begin{align}
\boldsymbol{A_{cl}}=
\begin{bmatrix}
0 & 1\\ 
-M & -2\sqrt{M}
\end{bmatrix}.
\end{align}
Then we can obtain a positive definite matrix $\boldsymbol{Q}$, where
\begin{align}
    \boldsymbol{Q}=\boldsymbol{A_{cl}^{\rm T}P}+\boldsymbol{PA_{cl}}=\begin{bmatrix}
1 &0 \\ 
0 &1 
\end{bmatrix}.
\end{align}
Hence, any trajectory in the local linear region ($\Omega_{6}$) will remain there indefinitely and once in $\Omega_{6}$, the trajectory converges to the target point.

In summary, the system in this paper with the control stategy \eqref{93-6}-\eqref{93-19} is globally stable. 

\rightline{Q.E.D.}
	
	\bibliographystyle{IEEEtranTIE}
	\bibliography{IEEEabrv,BIB_xx-TIE-xxxx}\ 

\end{document}